\documentclass[%
reprint, 
nobibnotes,
 amsmath,amssymb,
aps,
prb,
]{revtex4-1}

\usepackage{graphicx}
\usepackage{dcolumn}
\usepackage{enumitem}
\usepackage{bm}

\begin{document}

\newcommand\cutzts{Cu$_2$[TzTs]$_4$}

\newcommand{\V}[1]{\mathrm{#1}}



\title{Temperature Dependence of the Effective Interdimeric Exchange Interaction in a Weakly Coupled Antiferromagnetic Dimeric Copper Compound}

\author{Rafael Calvo}
 \affiliation{Departamento de F\'{i}sica, Facultad de Bioqu\'{i}mica y Ciencias Biol\'{o}gicas, Universidad Nacional del Litoral}
\affiliation{Instituto de F\'{i}sica del Litoral, CONICET-UNL, G\"{u}emes 3450, Santa Fe 3000, Argentina.}
 \email{calvo.rafael@conicet.gov.ar}
 
\author{Vinicius T. Santana}%
 \author{Otaciro R. Nascimento}
\affiliation{%
 Grupo de Biof\'{i}sica Molecular Sergio Mascarenhas, Departamento de F\'{i}sica e Ciencias Interdisciplinares, Instituto de F\'{i}sica de S\~{a}o Carlos, Universidade de S\~{a}o Paulo, CP 369, 13560-970 S\~{a}o Carlos, SP, Brazil}%




\date{\today}

\begin{abstract}
We report a variation with temperature ($T$) of the effective interdimeric interaction $J^\prime_{\V{eff}}$ in the antiferromagnetic (AFM) copper dimeric organic compound \cutzts~[N-thiazol-2-yl-toluenesulfonamidate Cu$^\V{II}$]. This $T$ dependence was obtained from measurements of the effects in the electron paramagnetic resonance (EPR) spectra of the proposed quantum phase transition associated to the exchange narrowing processes. \cutzts~contains exchange coupled pairs of Cu$^\V{II}$ spins $\bm{S_\V{A}}$ and $\bm{S_\V{B}}$ ($S$ = 1/2), with intradimeric AFM exchange coupling $J_0$ = (-115$\pm$1) cm$^{-1}$ ($\mathcal{H}_\V{ex} = -J_\V{0} \bm{S_\V{A}}\cdot \bm{S_\V{B}}$). The variation of the EPR line width of single crystals with field orientation around a ``magic angle'' where the transitions intersect, as well as the integrated signal intensity of the so-called ``U-peak'' of the powder spectrum were measured as a function of $T$. Modeling these data using arguments of exchange narrowing in the adiabatic regime considering the angular variation of the single crystal spectra and a geometric description, we find that the effective interdimeric coupling $|J^\prime_{\V{eff}}|$ associated with the exchange frequency $\omega_{ex}$ is negligible for $T<<|J_\V{0}/k_\V{B}|$, when the units are uncoupled, and $|J^\prime_{\V{eff}}|$ = (0.080 $\pm$ 0.005) cm$^{-1}$ ($|J^\prime_{\V{eff}}/J_0|$ = 7.0$\times$10$^{-4}$) at 298 K. Within this $T$-interval, two ranges of $|J^\prime_{\V{eff}}|$ with linear temperature variation but different slopes, with a kink at $\sim$80 K, are observed and discussed. This $T$-dependence arises from the growing population of the triplet state and its relevance in the properties of various arrays of DUs is discussed. Our experimental procedures and results are compared with those of previous works in ion radical salts and dimeric metal compounds. The relation between the effective coupling $|J^\prime_{\V{eff}}|$ and the real interdimeric exchange coupling $|J^\prime|$ related to the chemical paths connecting neighbor units, is discussed.

\end{abstract}

\maketitle


\section{\label{sec:level1}Introduction}

The seminal ideas about dimeric units (DUs)\cite{Bleaney1952,Abragam1970,Kahn1993} introduced $\sim$65 years ago at the sunrise of the magnetic resonance techniques have spread widely in interest and applications. A spin dimer ($\bm{S}_{i\bm{\V{A}}}$, $\bm{S}_{i\bm{\V{B}}}$) obeys the spin Hamiltonian:\cite{Abragam1970,Kahn1993,Bencini1990,Weil2007}
 \begin{eqnarray} \label{eq:ham} 
 \mathcal{H}_\V{0}(i) &=& \mu_\V{B} \bm{B_\V{0}} \cdot (\bm{g_\V{A}} \cdot \bm{S}_{i\bm{\V{A}}} + \bm{g_\V{B}} \cdot \bm{S}_{i\bm{\V{B}}} ) -J_\V{0} \bm{S}_{i\bm{\V{A}}}\cdot \bm{S}_{i\bm{\V{B}}} \nonumber \\ ~~ && + \bm{S}_{i\bm{\V{A}}} \cdot \mathcal{D}  \cdot \bm{S}_{i\bm{\V{B}}} ,
 \end{eqnarray}
\noindent where $ \bm{B_\V{0}} = \mu_\V{0} \bm{H} $ is the magnetic field and $\mu_\V{0}$ the vacuum permeability, $\bm{g_\V{A}}$ and $\bm{g_\V{B}}$ are the $\bm{g}$-matrices and $J_\V{0}$ is the isotropic intradimeric exchange coupling that, for antiferromagnetic interaction ($J_\V{0} < 0$) gives rise to a singlet ground state and an excited triplet state with energy {$|J_\V{0}|$}. $\mathcal{D}$ is the anisotropic spin-spin interaction matrix arising from dipole-dipole and anisotropic exchange giving rise to the fine structure of the electron paramagnetic resonance (EPR) spectra, a very appropriate technique to study dimeric units.\cite{Bencini1990,Weil2007,Furrer2013} Hyperfine couplings between the spins $\bm{S}_{i\alpha}$ and the nuclear spins, and antisymmetric spin-spin couplings that may contribute for DUs without a center of symmetry\cite{Bencini1990} may be added to $\mathcal{H}_\V{0}(i)$, but are not needed for our present analysis. Isolated DUs described by Eq. \eqref{eq:ham} are zero-dimensional entities providing valuable model systems in physics,\cite{Kahn1993,Furrer2013,Gatteschi2006,Benelli2015} chemistry\cite{Ferrando-Soria2017} and biochemistry.\cite{Holm1996,Solomon2014} The total Hamiltonian $\mathcal{H}$ for dimer arrays is a sum $\mathcal{H}_\V{0} = \sum_{i} \mathcal{H}_\V{0}(i)$  of Eq. \eqref{eq:ham} over the dimeric units, plus the contribution $\mathcal{H}^\prime$ containing exchange couplings between spins in neighbor DUs,
\begin{equation} \label{eq:totalham} 
\mathcal{H} = \mathcal{H}_\V{0} + \mathcal{H}^\prime =  \sum_{i} \mathcal{H}_\V{0}(i) - \sum_{i \ne j, \alpha, \beta} J_{i\alpha,j \beta}^\prime \bm{S}_{i\alpha} \cdot \bm{S}_{j\beta}.
\end{equation}
\noindent $\mathcal{H}^\prime$ gives rise to triplet excitons\cite{Nordio1966,Sachdev1990} and a variety of quantum properties of higher dimensional magnetic systems.\cite{Zapf2014,Ruegg2003,Sachdev2008,Giamarchi2008} Arrays of interacting DUs have fascinating quantum and magnetic properties that attracted great attention in recent years.\cite{Furrer2013,Zapf2014} Molecular magnetism,\cite{Kahn1993,Furrer2013,Gatteschi2006,Benelli2015} spin ladders,\cite{Barnes1993,Giamarchi2003} and Bose-Einstein condensation in quantum magnets\cite{Zapf2014,Giamarchi2008,Oosawa1999,Nikuni2000} are flowering upgrowths with $\mathcal{H}_\V{0}$ and $\mathcal{H}^\prime$ of Eq. \eqref{eq:totalham} having different characteristics, wide ranges of magnitudes, and multiple roles. Since $\sim$1960 important lines of research about ion radical salts were followed in parallel to the work in metallic dimeric units, with many similar procedures and results.\cite{Nordio1966,Chesnut1961a,Chesnut1961} 

Exchange narrowing (EN) theory\cite{Gorter1947,Anderson1953,Anderson1954,Kubo1954} states that an EPR spectrum split and broadened by intramolecular interactions as the dipole-dipole coupling, may be narrowed by dynamical processes involving $\mathcal{H}^\prime$, interchanging randomly the states of the perturbation. The classical papers on EN treated the changes of the EPR spectra of a paramagnet produced by weak exchange interactions transforming isolated spins in 3D arrays. They use the adiabatic approximation, where the perturbation averaged out by the exchange is essentially diagonal, a condition giving maximum transparency to the theory, and propose Gaussian fluctuations for the random local interactions.\cite{Gorter1947,Anderson1953,Anderson1954,Kubo1954,Abragam1961} Their results promoted progresses in the understanding of spin diffusion and spin waves, and played important roles in the progress of non-equilibrium statistical mechanics.\cite{Kubo1985} The basic processes of narrowing and merging the structures of the EPR spectra are described using a characteristic exchange frequency $\omega_\V{ex}$ and a distance $\delta\omega$ between the peaks of the structure. In paramagnetic compounds, $\omega_\V{ex}$ is related to the essentially $T$-independent exchange couplings $J^\prime$ between neighbor spins that may be defined as $|J^\prime| \approx \hbar \omega_\V{ex}$, that depends on the chemical paths connecting the spins. In fact, there is not one interaction, but a distribution over the neighbors. Since the short ranged exchange couplings decrease exponentially with distance, average values of $|J^\prime|$ are estimated from single crystal EPR measurements.\cite{Martino1995,Martino1996,Costa-Filho1999} Spectral changes are analyzed in three regimes:\cite{Anderson1954,Abragam1961}
\begin{enumerate}[label=(\alph*)]
\setlength{\itemsep}{-5pt}       
\item For $\omega_\V{ex} < \delta\omega$ the resonances broaden for increasing $\omega_\V{ex}$, blurring their structures and changing shapes.
\item For $\omega_\V{ex} > \delta\omega$ the resonances narrow for increasing $\omega_\V{ex}$, when the line structure merges to a single line.
\item Between the regimes (a) and (b), for $\omega_\V{ex} \sim \delta\omega$ abrupt quantum transitions are observed.
\end{enumerate}
The result\cite{Anderson1953} for the resonance width $\Delta\omega$ which has been applied to many situations is:
\begin{equation} \label{eq:omegarelation} 
\Delta \omega \approx \frac{\left(\delta \omega \right)^{2} }{\omega _{{\V{ex}}} }.  
\end{equation}
The merging of the resonances and their widths in the slow (a) and fast (b) regimes may be analyzed with a procedure proposed by Anderson\cite{Weil2007,Anderson1954,Abragam1961} where the resonance peaks are described by a complex line shape involving the unperturbed resonance frequencies and the exchange frequency. This method, equivalent to Bloch equations modified in the presence of exchange,\cite{Weil2007} has been used for chemical exchange processes in liquids,\cite{Kivelson1957} spin exchange in ion radical salts,\cite{Chesnut1961} and also small exchange couplings in solids.\cite{Martino1995,Hoffmann1983,Hoffmann1988,Hoffmann1994} EPR measurements in the regime (c) have advantages on accuracy and quantum intuitiveness and are used here as described below. 

If the material is composed of dimeric (or polymeric) antiferromagnetic (AFM) units instead of single spins, the EN phenomenon displays novel and interesting properties.\cite{Calvo2011a,Napolitano2008,Perec2010} We show below that the exchange frequency $\omega_{ex}$ in coupled dimeric arrays is related to an effective $|J^\prime_{\V{eff}}|$ and not to $J^\prime$ as in monomeric systems, a condition introducing severe changes in the exchange narrowing processes. Since the magnetic moment of each unit varies with $T$ approaching zero for $T<<|J_\V{0}/k_\V{B}|$, the effective coupling $|J^\prime_{\V{eff}}|$ should vary with the population of the excited triplet state and the units may become magnetically isolated in the lattice at low $T$. This $T$-dependence of the effective interaction between AFM DUs and its consequences on the spectra and properties of the dimer array is the goal of this investigation where we evaluate a $T$ dependent $|J^\prime_{\V{eff}}|= \hbar \omega_{ex}$, measuring the effects of the quantum phase transition occurring in the range (c). We study the merging and narrowing of the two allowed EPR absorptions within the excited triplet state around their intersection (crossing field $B_\V{U}$) arising from the anisotropic $\mathcal{D}$-term of Eq.  \eqref{eq:ham}, as a function of magnetic field orientation, moving through the quantum transitions occurring when the distance between these peaks equals the interaction $|J^\prime_{\V{eff}}|$.

We also consider the ``U-peak'', a unique feature of the powder spectra of weakly coupled DUs absent in single crystals, arising from the accumulation of EPR signal around $B_\V{U}$, from a range of field orientations where the resonances merge.\cite{Calvo2011a,Perec2010} Our method replaces the study of the quantum transition occurring when the lowest Zeeman component of the triplet state crosses the singlet state as a function of the magnitude of the magnetic field, used by researchers studying Bose-Einstein condensation,\cite{Zapf2014,Ruegg2003,Sachdev2008,Giamarchi2008,Oosawa1999,Nikuni2000,Tachiki1970} by one where the two allowed EPR transitions intersect as a function of the orientation of $ \bm{B_\V{0}}$. This replaces the method based in the Bloch equations used by other authors,\cite{Chesnut1961,Martino1995,Hoffmann1983,Hoffmann1988,Jones1963,Hoffmann1985a} avoiding less accurate fittings of line shapes.

We study the AFM compound \cutzts~ [N-thiazol-2-yl-toluenesulfonamidate Cu$^\V{II}$], having weakly coupled dimeric units\cite{Napolitano2008,Cabaleiro2008} ($|J^\prime/J_0| <$ 10$^{-3}$) and EPR is best suited for our purpose. $|J^\prime_{\V{eff}}|$ is evaluated as a function of $T$ with equal results from two independent sets of EPR data: line width as a function of field orientation in single crystals, and intensity of the ``U-peak'' in powder samples as a function of $T$. In order to achieve maximum transparency in the application of the EN theory to our data, we report measurements at $\sim$34 GHz, when non-diagonal contributions of the $\mathcal{D}$-term are small compared with the dominant Zeeman interaction.

\section{\label{sec:level2}Experimental details and results}

\begin{figure}
\includegraphics{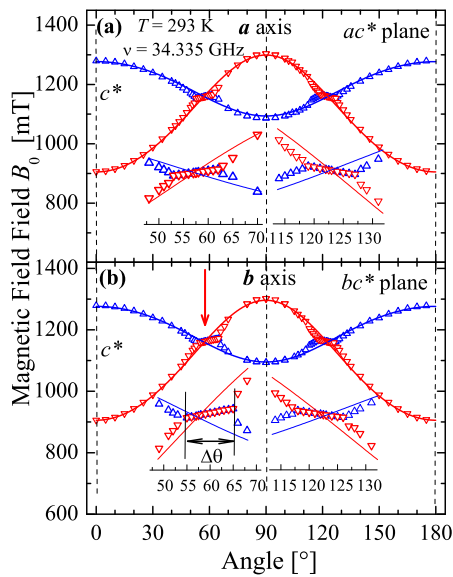}
\caption{\label{fig:ang_var} Angular variation of the resonances $M = \pm 1 \leftrightarrow 0 $ in the $ac^*$ (a) and $bc^*$ (b) crystal planes ($\bm{c^*} = \bm{a}\times\bm{b}$). Symbols are experimental results and solid lines are obtained fitting the data outside of the merged ranges. Insets display the merging around the magic angles. Arrows in (b) indicate the magic angle and the angular range $\Delta\theta$ where the line width measurements as a function of $T$ were performed.} 
\end{figure}

\begin{figure}
\includegraphics[scale=0.75]{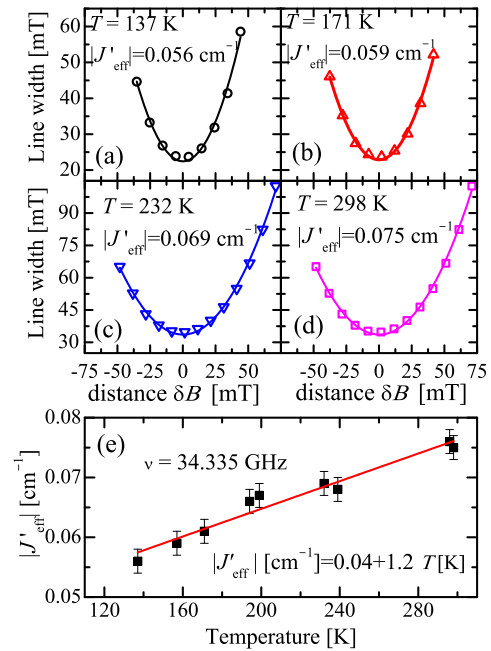}
\caption{\label{fig:paraboles} (a-d) Line widths of the merged signals around the magic angle in the $bc^*$ plane (arrow in Fig. \ref{fig:ang_var}b) at selected $T$, as a function of the distance between the fine structure peaks without merging. $|J^\prime_{\V{eff}}|$ is obtained from fits of Eq. \eqref{eq:omegarelation} to the data (solid lines). (e) Temperature variation of $|J^\prime_{\V{eff}}|$; symbols are experimental values with their estimated uncertainties. The line is a linear fit of the $T$-dependence of $|J^\prime_{\V{eff}}|$.} 
\end{figure}

\begin{figure}
\includegraphics[scale=0.6]{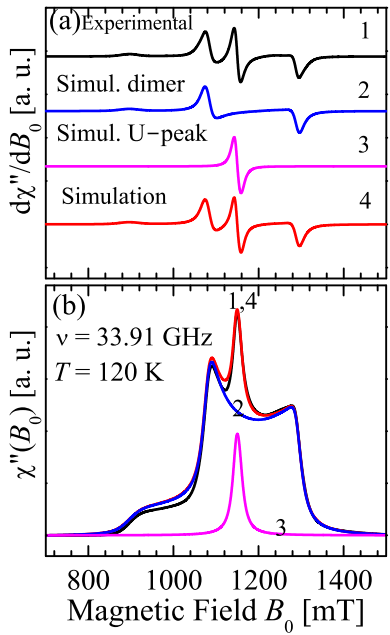}
\caption{\label{fig:spectra} (a1) EPR spectrum $d\chi''/dB_\V{0}$ at 120 K for a powder sample of \cutzts. (a2) Simulation obtained fitting Eq. \eqref{eq:ham} to the experimental result (it does not reproduce the central “U-peak”). (a3) Simulation of the U-peak. (a4) Sum of (a2) and (a3). (b) Integrals $\chi''(B0)$ of (a1-a4).} 
\end{figure}

The preparation, structure, and properties of dimeric \cutzts~and the used EPR techniques were described elsewhere.\cite{Calvo2011a,Napolitano2008,Cabaleiro2008} We collected EPR spectra of single crystals between 120 and 298 K at $\sim$34.3 GHz with a Varian E110 EPR spectrometer equipped with a nitrogen gas-flux $T$ controller, and in powder samples between 4 and 298 K, using a Bruker 500 spectrometer working at $\sim$33.9 GHz with a helium gas $T$-controller. Matlab\cite{Matlab2015} and Easyspin\cite{Stoll2006} 5.1.9 were used in the spectral calculations and fits. The parameters of the spin Hamiltonian $\mathcal{H}_\V{0}(i)$ for \cutzts~reported previously\cite{Napolitano2008} were verified, taking into account that the principal values $D$ and $E$ of the $\mathcal{D}$-matrix\cite{Weil2007} for two interacting 1/2-spins of Eq. \eqref{eq:ham} are twice of those for the spin $S$ = 1 model used before.\cite{Napolitano2008} Besides, the rhombic contribution $|E|$ is much smaller than the axial contribution $|D|$, and may be discarded, and the $\bm{g}$-matrices $\bm{g_\V{A}}$ and $\bm{g_\V{B}}$ are considered equal, with $g_{\V{x}}$ and $g_{\V{y}}$ differing within the experimental uncertainties. This allows assuming axial symmetry for the problem with $g_{//}$ = 2.232, $g_{\bot}$ = 2.045 and $|D|$ = 0.390 cm$^{-1}$ and produces an anisotropy $\delta\omega \propto |D|(3\cos^2\theta - 1 )$ of the line distance, where $\theta$ is the angle between the magnetic field and the axial symmetry direction. Figure \ref{fig:ang_var} displays the angular variation of the positions of the two EPR transitions $M = \pm 1 \leftrightarrow 0 $ of the excited $S$ = 1 spin triplet at 34.34 GHz and 293 K in the $ac^*$ and $bc^*$ planes of a single crystal sample of \cutzts~and simulations obtained with these parameters. 
The distance $\delta\omega$ between these peaks varies due to the $\mathcal{D}$-term in Eq. \eqref{eq:ham}, and their positions cross ($\delta\omega$ = 0) at the so-called magic angles in cones at 54.7${}^\circ$ and 125.3${}^\circ$ around the $\bm{z}$-axis (where $3\cos^2\theta - 1 = 0$).\cite{Weil2007} Insets of Fig. \ref{fig:ang_var} enhance the angular ranges where $\delta\omega \le \omega_\V{ex}$, where the two peaks merge due to the quantum entanglement produced by the interdimeric couplings. Associated to this collapse, the resonance displays a strong narrowing, that was used to calculate $|J^\prime_{\V{eff}}|$ at nine values of $T$ in the range 120 $< T <$ 298 K in the merged region around the magic angle $\theta$ = 54.7${}^{\circ}$ in the $bc^*$ plane (red arrow in Fig. \ref{fig:ang_var}b), using Eq. \eqref{eq:omegarelation}. Figures \ref{fig:paraboles}a-d display the widths of the merged resonances as a function of $\delta\omega$ between the resonances in the absence of merging calculated from the fit shown in Fig. \ref{fig:ang_var}, and the values of $|J^\prime_{\V{eff}}|=\hbar \omega_{ex}$ are collected in Fig. \ref{fig:paraboles}e, where a linear fit is included as a guide to the eyes. We also collected powder spectra at $\nu$ = 33.912 GHz between 4 and 298 K. Simulated spectra at each $T$ reproduce well the experimental result, with the exception of the central U-peak,\cite{Perec2010} not predicted by $\mathcal{H}_\V{0}(i)$ of Eq. \eqref{eq:ham}. Thus, we approximated this peak as arising from a single spin 1/2 that, when summed to the simulated dimeric spectrum, reproduces the full measured spectra. As an example of the analyses performed at each $T$, Figs. \ref{fig:spectra}a,b display the spectra $d\chi''/dB_\V{0}$ and the integrated $\chi''(B_\V{0})$, respectively, at $T$ = 120 K; a1 shows the observed spectrum, a2 and a3 are the simulations obtained with Eq. \eqref{eq:ham} and that for the U-peak; a4, sum of a2 and a3, is in good agreement with a1. Equal results for $\chi''(B_\V{0})$ are shown in Figs. \ref{fig:spectra}b 1-4.

\section{\label{sec:level3}Analysis of the data}

\begin{figure}
\includegraphics[scale=0.75]{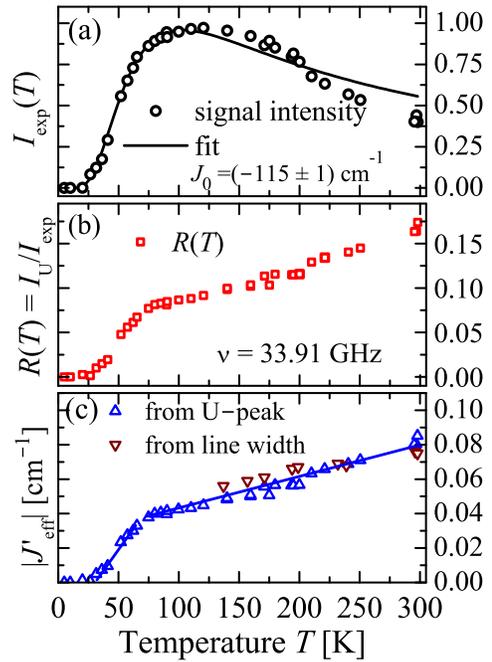}
\caption{\label{fig:joft} (a) Integrated area of $d\chi''/dB_\V{0}$ in (Fig. \ref{fig:spectra} a1) as a function of $T$. The line is a fit to Eq. \eqref{eq:bb} normalized to a maximum of 1, providing $J_0$. (b) Ratio $R(T)$ between the areas of the U-peak and of the full spectrum. (c) Up and down triangles display $|J^\prime_{\V{eff}}|$ calculated from the intensity of the U-peak and from the line width, respectively. Solid lines emphasize the different behavior of $|J^\prime_{\V{eff}}|$ in the low and high $T$ ranges.} 
\end{figure}

The intensity $I_\V{exp}(T)$ obtained by double integration of the powder spectra and the ratio $R$ between the integrated intensity $I_\V{U}(T)$ of the U-peak and $I_\V{exp}(T)$ are plotted in Figs. \ref{fig:joft}a and \ref{fig:joft}b. The Bleaney and Bowers equation:\cite{Bleaney1952,Kahn1993}
\begin{equation} \label{eq:bb} 
I_\V{exp}(T) \propto \frac{1}{T[3+\exp(-J_\V{0}/k_\V{B}T)]},  
\end{equation} 
\noindent normalized to a maximum value of one was fitted to the observed $I_\V{exp}(T)$, obtaining for the intradimer exchange coupling $J_\V{0}$ = (-115$\pm$1) cm$^{-1}$, similar and more accurate than the value reported before\cite{Perec2010} because of the wider $T$ range of the data. The U-peak becomes stronger at high $T$ in nearly axially symmetric arrays of DUs because it collects in a narrow magnetic field range the response of all units with $\bm{B}_\V{0}$ oriented near the magic angle where the two peaks are collapsed. In addition, for these field orientations the signal is narrowest (Fig. \ref{fig:paraboles}), and consequently larger. Since ambiguities are found in the literature, we mention that the U-peak behaves different from signals arising from paramagnetic monomeric copper contaminants whose EPR responses grow in intensity with decreasing $T$ (see \textit{e.g}., Sartoris \textit{et al}. \cite{Sartoris2015} and \v{S}im\'{e}nas \textit{et al}. \cite{Simenas2015b}), while the intensity of the U-peak decreases with decreasing $T$ and disappears at low $T$.
The fraction $R$ of DUs in an angular range $\pm \Delta\theta /2$ around the magic angle (Fig. \ref{fig:ang_var}b) is,
\begin{equation} \label{eq:r}
R = \frac{\int_{\theta_\V{M}-\Delta\theta/2}^{\theta_\V{M}+\Delta\theta/2} \sin{\theta}d\theta}{\int_{0}^{\pi/2} \sin{\theta}d\theta} = 2 \sin{\theta_\V{M}}\sin{(\Delta\theta /2)}.
\end{equation} 
So, $\Delta\theta = 2\sin^{-1}[R/(2\sin{\theta_\V{M}})]$ ($\cong R/\sin{\theta_\V{M}}$ for small $R$). We also calculated $\Delta\theta$ in terms of $|J^\prime_{\V{eff}}|$ from the angular variation of the positions of the EPR lines obtained fitting Eq. \eqref{eq:ham} to the data in Fig. \ref{fig:ang_var}, considering that the peaks $M = \pm$1$\leftrightarrow$0 are merged when their distance $\delta\omega \le \omega_\V{ex}$. With this condition we find that the effective exchange coupling $|J^\prime_{\V{eff}}|=\hbar \omega_{ex}$ for \cutzts~is linearly related to $R$ as $|J^\prime_{\V{eff}}(T)|$[cm$^{-1}$] = 0.49$R(T)$, and measurement of $R$ from the powder spectra allow evaluating $|J^\prime_{\V{eff}}|$ in the $T$-range 4-298 K, wider than the range of the line width measurements of Fig. {\ref{fig:paraboles}}. The linear relation between $|J^\prime_{\V{eff}}|$ and \textit{R} is maintained while $\Delta\theta$ is small and $\sin{(\Delta\theta /2)} \approx \Delta\theta /2$ in Eq. \eqref{eq:r}. The values of $|J^\prime_{\V{eff}}|$ are shown in Fig. \ref{fig:joft}c together with those obtained from the line width measurements in a narrower $T$-range (Fig. \ref{fig:paraboles}); the agreement of the two data sets is excellent considering the simplicity of the analysis. Our results for \cutzts~indicate that $J^\prime_{\V{eff}} \sim$0 below $\sim$25 K, and the dimeric units become uncoupled, and it increases linearly with $T$ above 25 K up to a kink at 80 K, where the slope decreases to $\sim$half and stays constant up to 298 K, when $|J^\prime_{\V{eff}}| \sim$ 0.08 cm$^{1}$ and $|J^\prime_{\V{eff}}/J_0|$ = 7.0$\times$10$^{-4}$.

\section{\label{sec:level4}Discussion and Conclusions}

Using ideas of exchange narrowing (Eq. \eqref{eq:omegarelation}), we calculated between 100 and 298 K a $T$-dependence of the effective interdimeric exchange interaction $|J^\prime_{\V{eff}}|$ for the antiferromagnetic compound \cutzts~from EPR line width measurements in single crystals with $\bm{B}_\V{0}$ oriented in the neighborhood of the magic angle. In a wider $T$-range, 4-298 K, we obtained equal results from measurements of the relative intensity of the so-called U-peak of the powder spectrum. In a previous work,\cite{Perec2010} we proposed that this peak is a consequence of the interdimer coupling $|J^\prime_{\V{eff}}|$.\cite{Calvo2011a,Perec2010} Here we prove that its relative signal intensity $R$ (fraction of the sphere of field orientations in the powder sample where the condition $\delta\omega \le \omega_\V{ex}$ holds) allows evaluating $|J^\prime_{\V{eff}}|$ (exchange narrowing phenomena do not depend on the sign of the interaction). The observed $T$-dependence is a consequence of the depopulation with decreasing $T$ of the excited triplet state, and the effective value of $J^\prime_{\V{eff}}$ becomes zero when the average magnetic moment of the AFM DUs cancels out. Our results explain why dimeric materials may not show the characteristic fine structure\cite{Weil2007} of the dimeric EPR spectra for interacting DUs, even when the expected magnitude of the dipole-dipole interaction would suggest such structure. They also explain changes with $T$ of the fine and hyperfine structures of DUs, as in the results reported by Lancaster \textit{et al}.,\cite{Lancaster2014} Sartoris \textit{et al}.,\cite{Sartoris2015} and Khadir \textit{et al}.\cite{Khadir2015a} for organic dimeric Cu$^{\V{II}}$ compounds, where no fine structure, at the $T$ of the experiments, or U-peaks in the powder spectra are observed, because the collapse of the fine structure occurs for all orientations of $\bm{B}_\V{0}$ (because $|J^\prime_{\V{eff}}| > |D|$), and not in restricted angular ranges. The usefulness of the U-peak to determine interdimeric interactions is limited to cases where $|J^\prime_{\V{eff}}|$ significantly less than $|D|$. In other cases this peak would be wide and thus weak and barely observable. Our results for the variation with $T$ of $J^\prime_{\V{eff}}$ also explain the type of hyperfine structure observed in dimeric compounds. Sebastian \textit{et al}.\cite{Sebastian2006} reported that, at 4.5 K, the spectrum of dimeric BaCuSi$_{2}$O$_{6}$ displays hyperfine structure typical of monomeric Cu$^{\V{II}}$ (present as an impurity trace), that merges and disappears at 9 K, because the contribution of the spectrum of monomeric Cu$^{\V{II}}$ contaminant decreases with increasing $T$ as 1/$T$. Dimeric hyperfine structure was not observed in that work, where it should be wiped out by the interdimeric exchange, because $|J^\prime_{\V{eff}}|$ is greater than the hyperfine coupling parameter $|A|$. Instead, the dimeric Cu$^{\V{II}}$ compound with a pyrophosphate tetra-anion and 2,2'-bipyridylamine\cite{Sartoris2015} displays a rich EPR spectral variation with $T$, that we now associate to the $T$-dependence of $J^\prime_{\V{eff}}$. Hyperfine structure characteristic of dimeric units is observed at 12 K, but disappears at higher $T$ because of the increase of $|J^\prime_{\V{eff}}|$, and at lower $T$, because the amplitude of the dimeric signal becomes negligible compared with that of paramagnetic Cu$^{\V{II}}$ contaminants, and display monomeric hyperfine coupling at 4.5 K, as in the study of BaCuSi$_{2}$O$_{6}$.\cite{Sebastian2005} Zvyagin \textit{et al}.{\cite{Zvyagin2006}} reported the split of the EPR spectrum of BaCuSi$_{2}$O$_{6}$ along a crystal axis below $T \sim$ 9 K, showing a characteristic dimeric behavior with the two $M = \pm 1 \leftrightarrow 0 $ transitions and the forbidden peak at half field. Considering our present results and their Figs. 3 and 4, we calculate $|J^\prime_{\V{eff}}| \sim 3|D|/2 \sim $ 0.15 cm$^{-1}$ from the collapse of the two peaks at $T \sim$ 10 K. We attribute this collapse to the increase of $|J^\prime_{\V{eff}}|$ with increasing $T$, as a consequence of $\omega_{ex}$  becoming relevant compared to $D$ and consequently, to $\delta \omega$.   This value of $|J^\prime_{\V{eff}}|$ for BaCuSi$_{2}$O$_{6}$ is $\sim$ one order of magnitude smaller than the value $J^\prime$ = 2.2 K estimated by Sasago \textit{et al}.{\cite{Sasago1997}}  This is coherent with our results, considering that 10 K is one fifth of the singlet-triplet splitting,{\cite{Zvyagin2006,Sasago1997}} and the population of the triplet state is small.


Hoffmann \textit{et al}.{\cite{Hoffmann1985a,Hoffmann1985}} studied structural dimeric copper compounds, whose dimeric fine EPR structure ($D$-term) is not resolved. The two peaks observed for most orientations of $\bm{B}_\V{0}$ arise from anisotropic $\bm{g}$-matrices of magnetically rotated Cu$^{\V{II}}$ ions in the lattices. The value of $|J^\prime|$ obtained in their works corresponds to the interaction between chemically identical but rotated Cu$^{\V{II}}$ ions, whose $g$-factors are averaged-out. The strong decreases of $|J^\prime|$ with increasing $T$ (a result opposite to ours) was attributed to the lattice vibrations, a process different than that observed by us in \cutzts.

Jones and Chesnut\cite{Jones1963} studied the effect of interdimeric exchange in various ion radical salts fitting the EPR line shapes to the predictions of Bloch equations for exchange coupled spin pairs.\cite{Weil2007} Their results for the temperature dependence of the interaction indicated the existence of an activated process with activation energies which, according to the authors, depend on the assumptions made in the fitting processes. They attribute the observed increase of the interaction with increasing $T$ to the varying population of the triplet state, as for the case of \cutzts~studied here. In view of their results we tried without success fitting $|J^\prime_{\V{eff}}(T)|$ as an activated process $J^\prime_{\V{eff}}(T) = J^\prime(0) \exp(-\Delta E / k_{\V{B}} T)$, where $J^\prime(0)$ is the limiting value for high $T$.

The observed $T$-dependence of the interdimeric interactions $|J^\prime_{\V{eff}}|$ observed in \cutzts~indicate that the environment of a DU acts as a single interaction like in effective field theories suggesting similar behavior for weakly interacting AFM metal clusters. Naively, we may assume that $|J^\prime_{\V{eff}}| \approx |J^\prime| \times$relative population of the triplet state, where $J^\prime$ is the actual average interaction between spins in neighbor units (Eq. {\eqref{eq:totalham}}). However, this assumption does not explain the change of the slope of the temperature variation at $T \sim$ 75 K. Possible explanations may require considering with more detail the dynamics of the spin excitations in \cutzts~using complementary experimental techniques. In any case we propose that $J^\prime \sim$ 0.08 cm$^{-1}$, as it is $|J^\prime_{\V{eff}}|$ for the maximum population of the triplet state.

The value of $J^\prime_{\V{eff}}$ and its variation with $T$ is important in various fields that gained much interest lately, as molecular magnets,\cite{Kahn1993,Benelli2015} spin excitations in arrays of AFM clusters,\cite{Furrer2013,Zapf2014} quantum phase transitions,\cite{Sachdev2008,Sachdev2011a} quantum spin-ladders,\cite{Barnes1993,Bouillot2011,Cizmar2010,Schmidiger2013} Bose-Einstein condensation in quantum magnets,\cite{Zapf2014,Sachdev2008,Giamarchi2008,Nikuni2000,Kato1999} and in the study of phase transitions and thermodynamic behavior studied earlier by Tachiki \textit{et al.}\cite{Tachiki1970,Tachiki1970v46} in dimeric Cu(NO$_{3}$)$\cdot$2.5H$_{2}$O and more recently in other metal-organic materials.\cite{Lancaster2014,Goddard2012,Brambleby2017} 

 Studies\cite{Zapf2014,Oosawa1999,Nikuni2000,Ruegg2005} of statistical properties of systems where the interactions $|J^\prime_{\V{eff}}|$ between dimeric units are larger than in \cutzts use inelastic neutron scattering\cite{Ruegg2003,Schmidiger2013} and thermodynamic measurements\cite{Brambleby2017} providing information about the spin excitation bands arising from this interaction. These techniques are more complex than EPR to evaluate the $T$-dependence of $J^\prime_{\V{eff}}$. The actual ratio $|J^\prime_{\V{eff}}/J_{\V{0}}|$ may favor using one technique, or studying different phase transitions, as the crossing of a Zeeman level of the excited spin triplet with the ground singlet as a function of the field intensity, or the crossing of two EPR transitions within the excited triplet as a function of the field orientation, as exploited here. Even if some arguments are common to the two cases where quantum phase transitions are observed, much more theoretical work exists for the spin wave analysis of the singlet-triplet transition as a function of the field intensity,\cite{Zapf2014} than for the case of transitions within the spin triplet. In cases where $|J^\prime_{\V{eff}}/J_{\V{0}}|$ is small and the dimeric structure of the EPR spectrum is observed, as for \cutzts, the temperature dependence of the intensity $R$ of the U-peak provides a simple and accurate way to evaluate $|J^\prime_{\V{eff}}(T)|$ that may be related to the spin excitations on the material.

Along this work we consider that the exchange narrowing process producing the collapse of the structure gives rise to quantum phase transitions{\cite{Sachdev2011a}} in the range $\omega_{ex} \sim \delta \omega$ between the slow and fast fluctuation Anderson's{\cite{Anderson1953,Anderson1954}} regimes when the exchange frequency equals the splitting $\delta \omega$ between the collapsing peaks. We sweep through these transitions changing $|J^\prime_{\V{eff}}|$ with $T$, or $\delta \omega$ with the orientation of the magnetic field. The entanglement of the triplet state wave functions produced by small interdimeric interactions is responsible for important changes of the spectra. These interesting concepts{\cite{Pastawski2007}} require further experimental and theoretical work.

\begin{acknowledgments}
This investigation was supported by CAI+D-UNL in Argentina and by CNPq in Brazil. RC is a member of CONICET, Argentina. We acknowledge to GPOMS (Laborat\'{o}rio de Propriedades \'{O}pticas e Magn\'{e}ticas de S\'{o}lidos) at the University of Campinas, SP, Brazil, for the use of the Q-band Bruker EPR spectrometer. We thank Dr. Jes\'us Castro (Departamento de Qu\'imica Inorg\'anica, Universidade de Vigo, Galicia, Spain) for preparing and kindly  supplying the samples used in this investigation.

We respectfully dedicate this paper to the memory of Dr. Lia M. B. Napolitano who participated in earlier stages \cite{Napolitano2008,Cabaleiro2008} of this investigation while working in her PhD degree at the Universidade de S\~{a}o Paulo.

\end{acknowledgments}

\end{document}